\documentclass[preprint,10.5pt]{article}
\usepackage[utf8]{inputenc}
\usepackage[english]{babel}

\usepackage{caption}
 \usepackage[margin=1.5in,footskip=0.4in]{geometry}

\usepackage{graphicx, caption}
\usepackage{amsmath, amssymb, color}

\usepackage{natbib}

\usepackage{boxhandler}
\usepackage{natbib}

\usepackage{etoolbox}
\usepackage{natbib}
\newcounter{bibcount}
\makeatletter
\patchcmd{\@lbibitem}{\item[}{\item[\hfil\stepcounter{bibcount}{\thebibcount.}}{}{}
\renewcommand\NAT@bibsetup%
   [1]{\setlength{\leftmargin}{\bibhang}\setlength{\itemindent}{-\parindent}%
       \setlength{\itemsep}{\bibsep}\setlength{\parsep}{\z@}}
\makeatother
\definecolor{capri}{rgb}{0.0, 0.75, 1.0}

\usepackage{array}
\newcolumntype{P}[1]{>{\centering\arraybackslash}p{#1}}

\newcount\exno

\def\exercise{\medbreak\advance\exno by 1 \noindent
   {\bf\thechapter.\number\exno}\quad}

\def\Span{\mathop{\rm span}\nolimits}

\def\det{\mathop{\rm det}\nolimits}

\def\pc{\mathop{\rm pc}\nolimits}

\def\vfrac#1#2{{\textstyle{\frac{#1}{#2}}}}
\def\vhalf{{\textstyle{\frac12}}}
\def\Real{{\mathbb R}}
\def\G{{\cal G}}
\def\X{{\cal X}}

\def\one{{\mathbf 1}}
\usepackage{boxhandler}
\usepackage{natbib}

\begin{document}


\title{
A likelihood analysis of quantile-matching transformations \\
\small \ }

\author{
Peter McCullagh \\
University of Chicago\\
\\
Micol Federica Tresoldi \\
University of Chicago\\
}

\maketitle

\begin{abstract}
Quantile matching is a strictly monotone transformation that sends the observed response values $\{y_1, \dots, y_n\}$ 
to the quantiles of a given target distribution. 
A likelihood based criterion is developed for comparing one target distribution with another
in a linear-model setting.

   \end{abstract}
   \hspace{1cm}

\section{Introduction}
In applied statistical work, it is frequently necessary to transform the response variable prior to fitting a linear Gaussian model. 
This entails identifying a transformation $g\colon \Real\to\Real$ and applying it
component-wise to the vector $Y\in\Real^n$ in the hope that the transformed variable $gY$
might be approximately normally distributed $N_n(\mu, \Sigma)$. Here, $\mu $ belongs to $\X $, the image subspace of the model matrix $X$, which is of order $n\times p$ and known. The covariance matrix $\Sigma$ belongs to some family of covariance matrices $\Theta$. For instance $\Theta$ might be the convex cone generated by a set of given symmetric non-negative definite matrices $V_i$, each representing some known relationship among the observational units.

According to this scenario, the joint density of the observation $Y$ at the point $y\in\Real^n$ is
\begin{equation}\label{density}
(2\pi)^{-n/2} |\Sigma|^{-1/2} e^{-(gy - \mu)'\Sigma^{-1}(gy-\mu)/2} \prod_{i=1}^n |g'(y_i)|
\end{equation}
provided that $g$ is differentiable and invertible.
In practice, it is reasonable to consider only strictly monotone differentiable functions, or real diffeomorphisms.

Perhaps the most widely used transformation in applied work is the power transformation proposed by \cite{box1964analysis}. Provided that $\one \subset \X$ and all observations are strictly positive, the transformation may be taken in the form 
$y\mapsto (y^g - 1)/g$ for some scalar~$g$, with the limit $g \to 0$ corresponding to the log function. 
The profile log likelihood for $g$ is
\[
l_p(g) = -\vhalf \log  |\hat{\Sigma}_g| +(g-1) \sum_{i=1}^n \log(y_i)
\]
provided that the maximum-likelihood estimate $\hat\Sigma_g$ exists.
By plotting $l_p(g)$ against $g$ one can easily check whether there is
a clear maximum in the range of interest, which is typically $-1\le g \le 1$. 
The underlying logic, however, aims to find ``a metric in terms of which the finding may be succinctly expressed'' (Box and Cox, 1964). 
So in practice, the logarithm and the identity are the transformations which are ordinarily chosen, leaving the reciprocal, square root or cube root to cases where there is a reasonable justification based on the physical units of measurement. 

More recently, in certain \lq big-data\rq\ settings arising in a variety of genetic research studies, another kind of transformation has become popular.
The rank-based direct inverse normal transformation (INT), (\cite{mccaw2019omnibus}; \cite{beasley2009rank}; \cite{servin2007imputation})
transforms the observed values of $Y$ so that the marginal distribution is standard Gaussian. 
However, unless $\X = \one$ or all of the effects are negligible, this procedure does not guarantee that $gY \sim N_n (\mu, \Sigma)$. 

This paper proposes a likelihood-based criterion for comparing one rank-based transformation with another. 
These quantile-matching transforms are defined in such a way that the transformed values can be considered to be realizations of a random vector with each component being marginally distributed according to some target distribution $G$. An illustration for two simulated row and column designs is presented in section 4.1 and 4.2.

\section{Likelihood for Gaussian models}

To specify the likelihood function, it is necessary to identify the set of transformations $g\in\G$ under consideration,
plus the mean-value space $\X=\Span(X)$ and the space $\Theta$ of covariance matrices.
To be clear, these moment spaces are moment spaces for the transformed variable $gY$, not for~$Y$.
As a function on $\G\times\X\times\Theta$, this density is the likelihood function.

 It is helpful at this stage to insert two technical conditions.
 First, the space~$\one$ of constant $n$-vectors is a subspace of~$\X$;
 this is not required in the theory of linear models, but it is universal in applied work.
 Second, the space of covariance matrices is a cone, i.e.,~$\Sigma\in\Theta$ implies
$\tau\Sigma\in\Theta$ for every scalar multiple $\tau > 0$.
Both conditions are mathematically essential but relatively benign; 
the cone need not be convex.
The cone condition extends to $\Sigma^{-1}$ and ensures that the maximum-likelihood estimate 
$\hat\mu_g, \,\hat\Sigma_g$ for fixed~$g$ satisfies
\[
(gy - \hat\mu_g)'\hat \Sigma_g^{-1}(gy-\hat\mu_g) = n.
\]
As a consequence, the profile log likelihood for the transformation $g\in\G$ is
\begin{equation}\label{profile_llik}
l_p(g) = - \vhalf\log\det(\hat\Sigma_g) + \sum_{i=1}^n \log |g'(y_i)| .
\end{equation}
Finally, for all scalars $a, b\neq0$, the cone condition and $\one\subset\X$
imply $l_p(a + bg; y) = l_p(g; y)$,
so that the profile likelihood is invariant to affine composition.
In other words, the transformations $y\mapsto g(y)$ and $y\mapsto a+b g(y)$ are equivalent for this comparison:
$gY \sim N(\mu, \Sigma)$ implies $a + b gY \sim N(a+b\mu, b^2\Sigma)$, and vice-versa.

The preceding analysis assumes that the maximum-likelihood estimate $\hat\mu_g, \hat\Sigma_g$ exists.
Existence and uniqueness cannot be guaranteed in general, but failure is rare in practice
provided that $p < n$ and the residual space is adequate to estimate all variance components.

\section{Likelihood ratios}
\subsection{Quantile-matching transformation}
Denote by $F$ the one-dimensional marginal distribution function of the response $Y_i$, or the average
of these distributions if they are not the same, and suppose it is continuous and strictly monotone. 
For any continuous strictly monotone cumulative distribution function $G$, the function
\begin{equation}
h = G^{-1}\circ F
\end{equation}
associates with each quantile of $F$ the corresponding quantile of $G$, and transforms $Y\sim F$  into $hY \sim G$.
In practice, however, we do not know the marginal distribution of $Y$ so, instead of $F$, we consider an empirical version, 
the percentile function relative to the set of the observed values. 

For any finite subset $A\subset\Real$ containing $m$ points counted with multiplicity, 
the percentile function at $t\in A$ is
the average of the left and right limits 
of the empirical distribution function,
\[
\pc(t) =\vhalf \hat{F}_m(t^-) +\vhalf \hat{F}_m(t^+).
\]
If there are no ties in $A$, the percentile values are the numbers $(2i-1)/2m$ for $1\leq i\leq m$. This corresponds to choosing $c = 1/2$ for the INT in equation (4) of \cite{mccaw2019omnibus}, which was first developed by \cite{bliss1956rankit} under the name of `rankit'. However, the choice of $c$ is not critical in the definition of the rank transform.
For all other values $t\not\in A$, $\pc(t)$ is defined to be any strictly-monotone differentiable interpolant
satisfying $0 < \pc(t) < 1$. 
So, apart from the points in $A$, the values of the percentile function are unspecified. 
The $G$ quantile-matching transformation is then defined as 
\begin{equation}\label{FQQ}
 t \mapsto G^{-1}\bigl(\pc(t)\bigr).
\end{equation}
When applied component-wise to the vector~$y$, this transformation converts the observed values into specific quantiles $q_i^G$ of the target distribution $G$, preserving order.

\subsection{Connections with rank regression}


\cite{kruskal1965analysis} first proposed modifying the Box-Cox proposal by considering the space of all monotone transformations rather than only power transformations. Similar proposals based on rank marginal likelihood
were developed by 
    \cite{pettitt1982inference}, \cite{pettitt1987estimates}, \cite{doksum1987extension}, and \cite{cuzick1988rank}, among others. 
All these approaches assume additivity plus independent and identically distributed errors.
They focus on obtaining the maximum-likelihood estimate for the regression coefficients based on the observed ranks, 
treating the transformation as nuisance parameter.
In order to avoid the numerically difficult integration step for the rank likelihood,  \cite{cuzick1988rank} proposed using
\begin{equation}\label{Cuzick}
z_i = F_\mu^{-1}(\hat{F}(y_i))
\end{equation}
in place of the expectation given the rank vector.
Here $\hat{F}$ is a modified version of the empirical distribution function, and $F_\mu(t)$ is the average distribution function of $gY_i$ under the assumption that $gY_i = \mu_i +\epsilon_i$, with $\epsilon_i \sim F_0$ and $F_0$ is known. So $F_\mu$ depends on 
the unknown fixed vector~$\mu$. The estimated transformation is then found, as a byproduct of the coefficients estimation, by substituting in \eqref{Cuzick} the maximum likelihood estimate of $\mu$ and interpolating between the data points. 

The focus of this paper is not so much on the computation of regression coefficients, but on the use of the
likelihood function to compare one proposed transformation with another.
For this purpose, the rank likelihood is uninformative because it is independent of the transformation.
We focus solely on quantile-matching transformations, as indexed by the target distribution~$G$.
The use of likelihood ratios for this class circumvents the problem of infinite likelihood,
and enables us to compare one target distribution directly with another.


\subsection{Likelihood ratios}
Let $g = G'$ be the density of the target distribution.
Since the derivative of the quantile-matching transformation is 
$\pc'(t) / g\bigl( G^{-1}\bigl(\pc(t)\bigr) \bigr)$,  the profile log~likelihood is 
\begin{equation}\label{Fqmllik}
l_p(G) = -\vhalf\log\det(\hat\Sigma_G) + \sum\log\pc'(y_i) - \sum\log  g\bigl( G^{-1}\bigl(\pc(y_i)\bigr) \bigr),
\end{equation}
where $\hat\Sigma_G$ is the maximum-likelihood estimate after transformation.
The last term appearing in \eqref{Fqmllik} is $n $ times the quadrature approximation to the entropy integral of $G$
\[
- \frac{1}{n}  \sum\log g (q^G_i ) = - \int \log g(x) \,d G(x) + O(n^{-1}).
\]

At first sight, \eqref{Fqmllik} appears to be unusable because it depends on the derivative of the percentile interpolant.
However, we can compare one target distribution $G$, with another, say $\tilde{G}$, since their log~likelihood ratio,
\begin{equation}\label{gglr}
-\vhalf\log\det(\hat\Sigma_{\tilde{G}} \hat\Sigma_G^{-1}) - \sum\log \tilde{g} \bigl( \tilde{G}^{-1}\bigl(\pc(y_i)\bigr)\bigr) +\sum\log g\bigl( G^{-1}\bigl(\pc(y_i)\bigr)\bigr),
\end{equation}
is unaffected by the interpolant.
If $n$ is sufficiently large, the two quadrature sums in \eqref{gglr} can be replaced with the corresponding integrals. 
The quadrature errors are typically $O(n^{-1})$ for both distributions, but if these are contiguous or similar, 
the quadrature error for the difference is $o(n^{-1})$.

As an example, suppose $G$ is the uniform distribution.
Its quantile-matching transformation $t \mapsto \pc(t)$ achieves a log likelihood
\begin{equation}\label{uqmllik}
-\vhalf\log\det(\hat\Sigma_U) + \sum\log\pc'(y_i). 
\end{equation}
For a direct comparison, consider the standard-Gaussian quantile-matching transformation 
\begin{equation}\label{GaussianQQ}
t \mapsto \Phi^{-1}\bigl(\pc(t)\bigr).
\end{equation}
The derivative is $\pc'(t) / \phi\bigl(\Phi^{-1}\bigl(\pc(t)\bigr)\bigr)$, which implies that the profile log likelihood is  
\begin{equation*}\label{gqmllik}
-\vhalf\log\det(\hat\Sigma_\Phi) + \sum\log\pc'(y_i) -  \sum\log\phi\bigl(\Phi^{-1}\bigl(\pc(y_i)\bigr)\bigr),
\end{equation*}
so the Gaussian-to-uniform log~likelihood ratio is 
\begin{equation}\label{gqlr}
-\vhalf\log\det(\hat\Sigma_\Phi \hat\Sigma_U^{-1}) + \vhalf \sum \bigl(\Phi^{-1}\bigl(\pc(y_i)\bigr)\bigr)^2 + \vfrac n 2 \log(2\pi).
\end{equation}
Since the Gaussian variance is 12 times that of the uniform, a first-order linear approximation suggests
$\hat\Sigma_\Phi \simeq 12 \hat\Sigma_U$, in which case the determinantal term in \eqref{gqlr} reduces to 
$-\vfrac n 2 \log (12) \simeq -1.242 n$.
If the response values are distinct, the sum of squared Gaussian quantiles satisfies $\sum \bigl(\Phi^{-1}\bigl(\pc(y_i)\bigr)\bigr)^2  = n + O(1)$, 
so the correction term in \eqref{gqlr} reduces to $\vfrac n 2 (1+\log(2\pi)) \simeq 1.419n$,
slightly over-compensating for the change of scale.

\subsection{Families of quantile transformations}

The Student-$t_\nu$ family includes the Cauchy distribution at $\nu=1$ and the Gaussian in the limit $\nu \to \infty$. 
The log likelihood ratio statistic for the comparison of $ t_{\nu}$ versus the Gaussian
is
\begin{equation}\label{llrgt}
-\vhalf\log\det(\hat\Sigma_\nu \hat\Sigma_\Phi^{-1}) - \sum\log f_{t_{\nu}} \bigl(t^{-1}_{\nu} \bigl(\pc(y_i)\bigr)\bigr)+ \sum\log\phi\bigl(\Phi^{-1}\bigl(\pc(y_i)\bigr)\bigr),
\end{equation}
where $ f_{t_{\nu}}$ denotes the Student~$t_\nu$ density function, and $\hat\Sigma_\nu $ is the 
maximum-likelihood estimate after transformation.

Alternatively, one can define a quantile-matching transform directly. 
For instance, the quantile function $q_{\alpha, \beta} (p) = p^\alpha\!/\alpha - (1-p)^\beta\!/\beta$ defines the family of transformations
\begin{equation}\label{galpha}
t \mapsto  \pc(t)^\alpha/\alpha -(1-\pc(t))^\beta/\beta,\, \hspace{.5cm}\, \alpha, \beta \in \mathbb{R}\,.
\end{equation}
Taking $\alpha = \beta$,  the log likelihood is
\begin{equation*}\label{uqmllik}
- \vhalf\log\det(\hat\Sigma_\alpha) + \sum\log\pc'(y_i) +  \sum \log (\pc(y_i)^{\alpha-1} + (1-\pc(y_i))^{\alpha-1})
\end{equation*}
since $q_\alpha (p) = F_\alpha^{-1}(p) = p^\alpha\!/\alpha - (1-p)^\alpha\!/\alpha$ implies that $f_\alpha(t) = (F_\alpha(t)^{\alpha-1} + (1-F_\alpha(t))^{\alpha-1})^{-1}$. 
The limit $\alpha \to 0$  corresponds to the logistic quantile-matching transformation 
\begin{equation}\label{logisticQQ}
t \mapsto \log\biggl( \frac{\pc(t)} {1 - \pc(t)} \biggr) .
\end{equation}
The derivative of this transformation is $\pc'(t) / \bigl(\pc(t)(1-\pc(t))\bigr)$,
so the logistic-to-uniform log~likelihood ratio is 
\begin{equation}
-\vhalf\log\det(\hat\Sigma_0 \hat\Sigma_U^{-1}) - \sum\log(\pc(y_i)(1-\pc(y_i))).
\end{equation}
The approximation $\sum\log\pc(y_i) \simeq \sum\log(1-\pc(y_i)) \simeq -n$
implies that the log likelihood ratio is approximately
\[
-\vhalf\log\det(\hat\Sigma_0\hat\Sigma_U^{-1})  +2n.
\]

It is worth emphasizing that the Gaussian regression model $Y\sim N_n(\mu, \Sigma)$ does not imply that
the $n$ components $Y_1,\ldots, Y_n$ have the same distribution,
nor does it imply that the histogram of $Y$-values should be close to Gaussian.
Unless $\X=\one$ or all of the effects are small, there is no compelling reason to expect that Gaussian quantile-matching 
should be more effective for present purposes than matching on other distributions,
even asymmetric distributions.
In most cases, however, Gaussian quantile-matching appears to be reasonably effective 
but not necessarily optimal.

\section{Simulated examples}
\subsection{Truth included}
As an illustration, we simulate data from a row-column design with independent and identically distributed
additive Gaussian row and column effects as follows:
\begin{verbatim}
nrows <- 50;  ncols <- 30;  n <- nrows*ncols
row <- gl(nrows, 1, n);  col <- gl(ncols, nrows, n)
set.seed(3142)
mu <- rnorm(nrows)[as.numeric(row)] + rnorm(ncols)[as.numeric(col)]
y <- 5 + mu + rnorm(n)
\end{verbatim}

When the response values are generated additively according to the Gaussian model,
the optimal transformation is the identity. Strictly speaking, the identity is not among the options accessible by quantile-matching as this latter is a function of the observed rank vector only. However, the identity can be closely approximated by choosing the quantile-matching transform corresponding to the true marginal distribution of $Y$. Thus, in this setting, $Y$ being normally distributed, the optimal quantile-matching transformation is given by the probit, i.e., the Gaussian quantile.
It therefore appears natural to consider the $t_\nu$ quantile-matching family 
in which the Gaussian is at the boundary $\nu = \infty$. 

We start by assuming the additive Gaussian model with subspace $\X=\hbox{\sl row+col}$ and covariance $\Sigma\propto I_n$. Disregarding the common term coming from the derivative of the percentile function, 
the profile log likelihood of the transformation as a function of $\nu$, is 
\begin{equation}\label{gtllik}
l_p(\nu) = -\frac{n}{2} \log(\hat\sigma^2_\nu) +  \sum\log f_{t_{\nu}} (t^{-1}_{\nu} (\pc(y_i)) ).
\end{equation}
The black line in Fig.~1 shows $l_p(\nu)$ as a function of $1/\nu$. 
As expected, the maximum is reached at or close to $1/\hat{\nu} = 0$.

Given the factorial design, one reasonable variation in the present setting is to use an additive Gaussian
random-effects model with $\X=\one$, and $\Sigma$ a linear combination of
the block matrices $I_n$, {\tt row} and {\tt col}.
For a balanced design such as this, maximum-likelihood estimates of all four parameters are available in closed form,
so the computations are not onerous.
The determinantal term in the profile log likelihood is now given by $-\vhalf \log \det(\hat{\sigma}^2_{R, \nu} {\tt row}  + \hat{\sigma}^2_{C, \nu} {\tt col} + \hat{\sigma}^2_{ \nu} I_n)$. As shown by the blue line in Fig.~1, $l_p(\nu)$ for the random effects model looks much the same as that for the fixed effects model,
except that all log likelihood values are reduced by approximately 140 units.
The reduction is not quite constant, but the maximum is still achieved at $1/\hat{\nu} = 0$.
\begin{figure}
\centerline{\includegraphics[scale = 0.75]{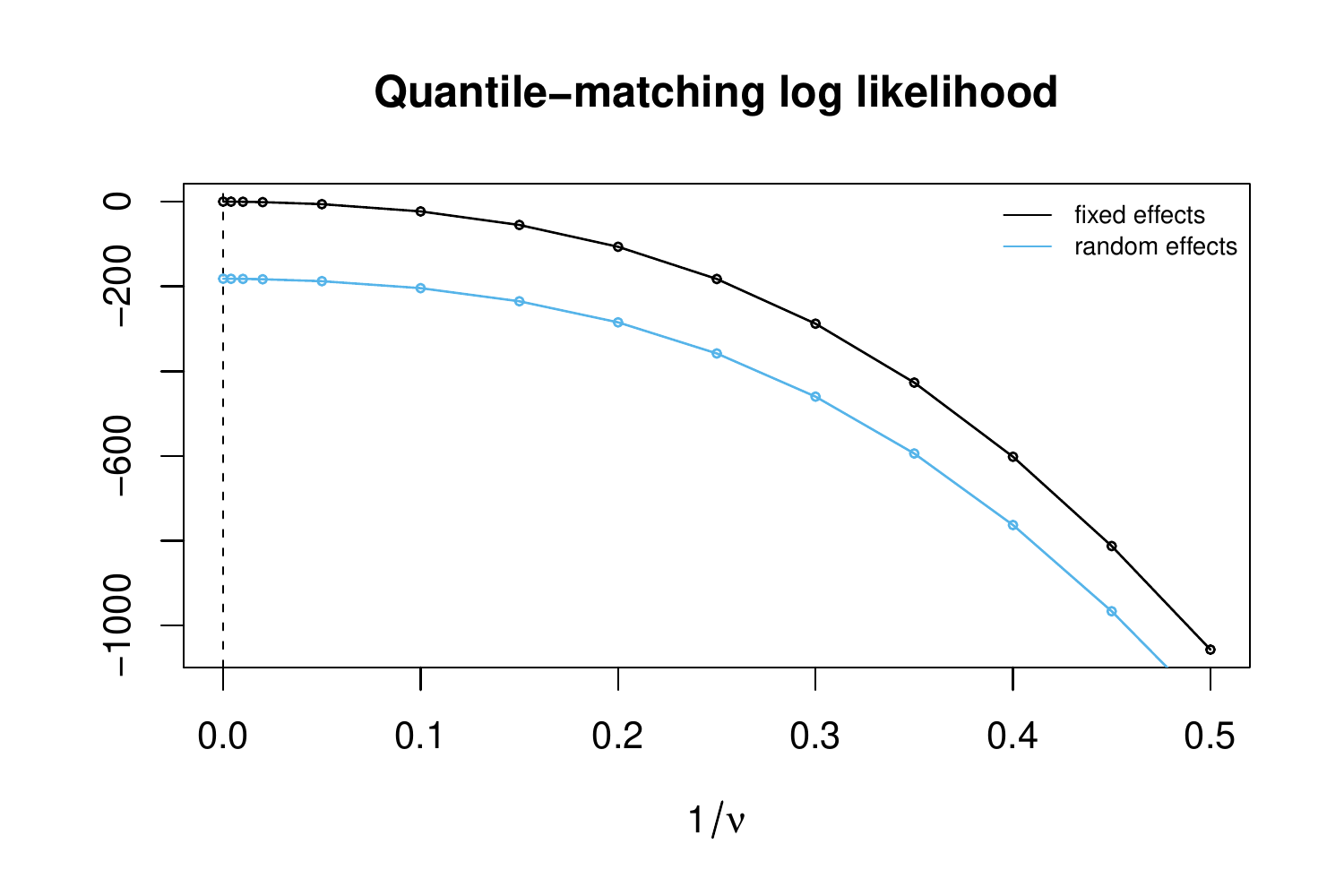}\hfil}
\caption{Profile log likelihood $l_p(\nu)$ plotted against $1/\nu$. The black and light blue lines refer to the fixed and random effects models, respectively. The dashed line indicates the maximum occurs at $1/\hat{\nu} = 0$, corresponding to the Gaussian quantile-matching transform.}
\end{figure}

\subsection{Truth not included}
We now repeat the same exercise with data not having normal marginal distribution. Specifically, we still simulate data from a row-column design but now the coefficients 
are independent Cauchy. The errors are $\epsilon_i \stackrel{i.i.d.}{\sim} N(0,1)$ which implies the distribution of $Y$ is symmetric but markedly non-Gaussian. As before, the optimal transformation for $y$ is the identity. In this case, however, we do not necessarily have among our options the quantile transform associated to the true marginal distribution of $Y$. 

We start by considering the family of transformations defined in  \eqref{galpha}, with $\alpha = \beta$, so that the limit $\alpha\to 0$ is the logistic model.
In practice, it suffices to focus on the range $-1 < \alpha < 1$, or some subset thereof.
Again, disregarding the common term coming from the derivative of the percentile function, 
we can compute 
the profile log likelihood corresponding to a given value of $\alpha$.
Fig.~2 shows $l_p(\alpha) $ plotted against $\alpha$. For these data, the maximum is reached at $\hat{\alpha} = -0.05$. 
If instead one considers the $t_\nu$-quantile family, the maximum occurs at $\hat{\nu} = 6.67$. The values of $l_p(\hat{\nu}) $ and $l_p(\Phi)$ are shown in Fig.~2.
Ordinarily, Gaussian quantile-matching is quite effective, but
for these data, the logistic quantile-matching function with $\alpha=0$ 
works appreciably better, comparable to the $t_{6.67} $ transform, the optimal among {the $t_\nu$-quantile family.

\begin{figure}
\centerline{\includegraphics[scale = 0.75]{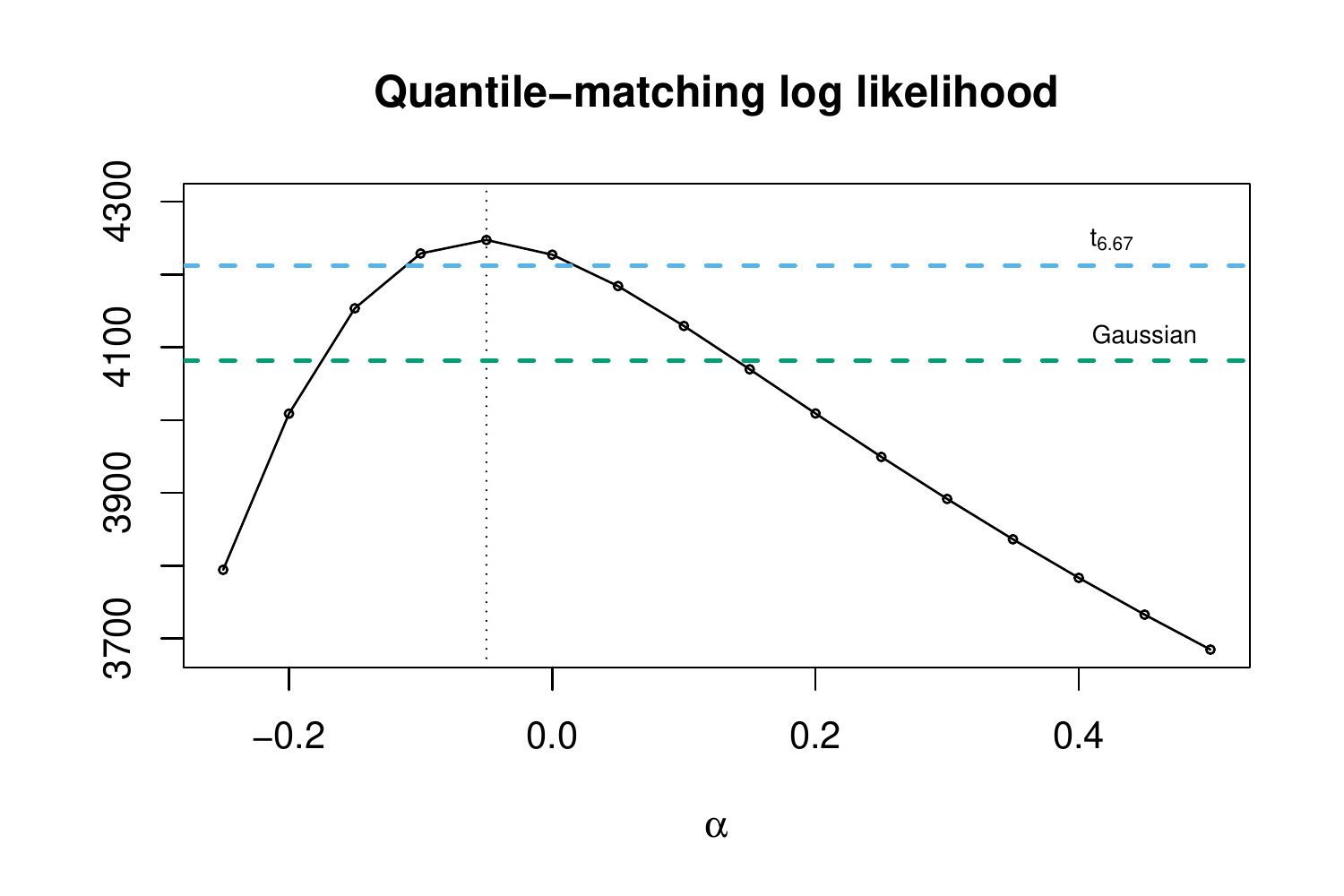}\hfil}
\caption{Profile log likelihood $l_p(\alpha)$. The dashed blue and green lines indicate the profile log likelihood for the $t_{6.67}$ and the Gaussian quantile-matching transforms, respectively. The dotted line indicates the maximum over the family \eqref{galpha} with $\alpha = \beta$, occurs at $\hat{\alpha} = -0.05$.}
\end{figure}

In this example, the quantile-matching families we considered did not include the quantile of the true distribution of $Y$. 
Nonetheless, among the transformations considered, the correlation matrix shows that the 
quantile transformation for $\hat\alpha = -0.05$ is maximally correlated with the optimum:
\[
\arraycolsep 10pt
\begin{array}{ccccc}
\multicolumn5{l}{\hbox{Correlations with quantile-transformed variables}\hss}\\
\hline \noalign{\smallskip}
 &\hat\alpha&   \hbox{logistic}   &\hbox{Gaussian} & t_{6.67}\\
\noalign{\smallskip}
\hbox{Identity} &  0.927 &0.917 &0.888 & 0.921 \\
\hline
\end{array}
\]

\bibliography{biblio}

@article{box1964analysis,
  title={An analysis of transformations},
  author={Box, George EP and Cox, David R},
  journal={Journal of the Royal Statistical Society: Series B (Methodological)},
  volume={26},
  number={2},
  pages={211--243},
  year={1964},
  publisher={Wiley Online Library}
}

@article{servin2007imputation,
  title={Imputation-based analysis of association studies: candidate regions and quantitative traits},
  author={Servin, Bertrand and Stephens, Matthew},
  journal={PLoS genetics},
  volume={3},
  number={7},
  pages={e114},
  year={2007},
  publisher={Public Library of Science}
}

@article{beasley2009rank,
  title={Rank-based inverse normal transformations are increasingly used, but are they merited?},
  author={Beasley, T Mark and Erickson, Stephen and Allison, David B},
  journal={Behavior genetics},
  volume={39},
  number={5},
  pages={580},
  year={2009},
  publisher={Springer}
}

@article{kruskal1965analysis,
  title={Analysis of factorial experiments by estimating monotone transformations of the data},
  author={Kruskal, Joseph B},
  journal={Journal of the Royal Statistical Society: Series B (Methodological)},
  volume={27},
  number={2},
  pages={251--263},
  year={1965},
  publisher={Wiley Online Library}
}

@article{pettitt1982inference,
  title={Inference for the linear model using a likelihood based on ranks},
  author={Pettitt, Anthony N},
  journal={Journal of the Royal Statistical Society: Series B (Methodological)},
  volume={44},
  number={2},
  pages={234--243},
  year={1982},
  publisher={Wiley Online Library}
}

@article{pettitt1987estimates,
  title={Estimates for a regression parameter using ranks},
  author={Pettitt, AN},
  journal={Journal of the Royal Statistical Society: Series B (Methodological)},
  volume={49},
  number={1},
  pages={58--67},
  year={1987},
  publisher={Wiley Online Library}
}

@article{cuzick1988rank,
  title={Rank regression},
  author={Cuzick, Jack},
  journal={The Annals of Statistics},
  volume={16},
  number={4},
  pages={1369--1389},
  year={1988},
  publisher={Institute of Mathematical Statistics}
}

@article{doksum1987extension,
  title={An extension of partial likelihood methods for proportional hazard models to general transformation models},
  author={Doksum, Kjell A},
  journal={The Annals of Statistics},
  volume={15},
  number={1},
  pages={325--345},
  year={1987},
  publisher={Institute of Mathematical Statistics}
}

@article{mccaw2019omnibus,
  title={Omnibus Inverse Normal Transformation Based Association Test Improves Power in Genome-Wide Association Studies of Quantitative Traits},
  author={McCaw, Zachary R and Lane, Jacqueline M and Saxena, Richa and Redline, Susan and Lin, Xihong},
  journal={bioRxiv},
  pages={635706},
  year={2019},
  publisher={Cold Spring Harbor Laboratory}
}

@article{bliss1956rankit,
  title={A rankit analysis of paired comparisons for measuring the effect of sprays on flavor},
  author={Bliss, CI and Greenwood, Mary L and White, Edna Sakamoto},
  journal={Biometrics},
  volume={12},
  number={4},
  pages={381--403},
  year={1956},
  publisher={JSTOR}
}

\end{document}